\documentclass[%
 aip,
rsi,%
 amsmath,amssymb,
reprint,%
]{revtex4-1}

\usepackage[version=3]{mhchem} 
\usepackage{graphicx}  
\usepackage{amsmath, amssymb, bm}
\usepackage{booktabs}
\usepackage{multirow}
\usepackage{color}
\usepackage{ulem}
\usepackage{array}

\begin{document}

\preprint{AIP/123-QED}

\title{Critical-point model dielectric function analysis of WO$_3$ thin films deposited by atomic layer deposition techniques}

\author{Ufuk K{\i}l{\i}\c{c}}
	\affiliation{Department of Electrical and Computer Engineering, University of Nebraska - Lincoln, Lincoln, Nebraska, USA}
    \affiliation{Center for Nanohybrid Functional Materials, University of Nebraska - Lincoln, Lincoln, Nebraska, USA}
\email{ufuk.kilic@huskers.unl.edu}
\homepage{http://ellipsometry.unl.edu}
\author{Derek Sekora}
	\affiliation{Department of Electrical and Computer Engineering, University of Nebraska - Lincoln, Lincoln, Nebraska, USA}
	\affiliation{Center for Nanohybrid Functional Materials, University of Nebraska - Lincoln, Lincoln, Nebraska, USA}
\author{Alyssa Mock}
	\affiliation{Department of Electrical and Computer Engineering, University of Nebraska - Lincoln, Lincoln, Nebraska, USA}
        \affiliation{Center for Nanohybrid Functional Materials, University of Nebraska - Lincoln, Lincoln, Nebraska, USA}
\author{Rafa\l{} Korlacki}
	\affiliation{Department of Electrical and Computer Engineering, University of Nebraska - Lincoln, Lincoln, Nebraska, USA}
        \affiliation{Center for Nanohybrid Functional Materials, University of Nebraska - Lincoln, Lincoln, Nebraska, USA}
\author{Elena M. Echeverr\'ia}
\affiliation{Department of Physics and Astronomy, University of Nebraska - Lincoln, Lincoln, Nebraska, USA}
\author{Natale Ianno}
	\affiliation{Department of Electrical and Computer Engineering, University of Nebraska - Lincoln, Lincoln, Nebraska, USA}
        \affiliation{Center for Nanohybrid Functional Materials, University of Nebraska - Lincoln, Lincoln, Nebraska, USA}
\author{Eva Schubert}
	\affiliation{Department of Electrical and Computer Engineering, University of Nebraska - Lincoln, Lincoln, Nebraska, USA}
        \affiliation{Center for Nanohybrid Functional Materials, University of Nebraska - Lincoln, Lincoln, Nebraska, USA}
    \author{Mathias Schubert}
	\affiliation{Department of Electrical and Computer Engineering, University of Nebraska - Lincoln, Lincoln, Nebraska, USA}
        \affiliation{Center for Nanohybrid Functional Materials, University of Nebraska - Lincoln, Lincoln, Nebraska, USA}
	\affiliation{Department of Physics, Chemistry, and Biology, IFM, Link\"oping, University, SE-58183 Link\"oping, Sweden}
    \affiliation{Leibniz Institute for Polymer Research, Dresden, Germany}

\date{\today}

\begin{abstract}
WO$_{3}$ thin films were grown by atomic layer deposition and spectroscopic ellipsometry data gathered in the photon energy range of 0.72-8.5 eV and from multiple samples was utilized to determine the frequency dependent complex-valued isotropic dielectric function for WO$_{3}$. We employ a critical-point model dielectric function analysis and determine a parameterized set of oscillators and compare the observed critical-point contributions with the vertical transition energy distribution found within the band structure of WO$_{3}$ calculated by density functional theory. We investigate surface roughness with atomic force microscopy and compare to ellipsometric determined effective roughness layer thickness.
\end{abstract}



\maketitle

\section{Introduction}

Transition-metal oxides such as tungsten tri-oxide (WO$_{3}$) continue to receive increasing interest due to their potential for use in photovoltaics,\cite{ou2013anodic} chemical gas sensing,\cite{zhuiykov2008morphology, kim2015acetone} electrochromic smart windows,\cite{lee2006crystalline} and optical switching\cite{granqvist2000electrochromic} applications, for example. Physical vapor deposition (PVD) processes such as sputtering\cite{sberveglieri1995wo3, depero1996structural, moulzolf2001stoichiometry} and thermal evaporation techniques\cite{lee2000fabrication, cantalini1996no2} and chemical vapor deposition (CVD) processes\cite{cross2003aerosol,chakrapani2016modulation} are convenient to fabricate WO$_{3}$ thin films. It has been reported that the electrical, optical, and photocatalytic properties of WO$_{3}$ thin films depend crucially on the growth conditions.\cite{zhuiykov2013atomically, liu2015few, labidi2005impedance, ping2013optical, liu2011water, zheng2015urea} For example, Subrahmanyam~\textit{et al.} examined the effects of the growth conditions during a sputtering process onto the optical and structural properties of WO$_{3}$ thin films.\cite{subrahmanyam2007optical} Hao~\textit{et al.} employed a spray pyrolysis method in order to fabricate WO$_{3}$ thin films, and investigated changes of their transient photoconductivity properties upon thermal annealing.\cite{hao2001transient}  Gullapalli~\textit{et al.} studied sputter-deposited nanocrystalline WO$_{3}$ films, and determined their transmittance and reflectance within the range of 1~eV--4.2~eV.\cite{gullapalli2010structural} Saenger~\textit{et al.} deposited amorphous WO$_{3}$ thin films by magnetron sputtering, and studied their polaron and phonon properties upon reversible electrochemical proton intercalation and reported the dielectric function in the spectral range of 0.037--3.34~eV.\cite{saenger2008polaron}. 
The dielectric function can provide insight into optical and electrical properties of a material\cite{xiong1993complex,li2014measurement} and has been investigated previously for WO$_3$ using several methods including ellipsometric techniques. For example, K. von Rottkay \textit{et al.} utilized ellipsometry and spectrophotometry to investigate the optical constants within the range 0.49~eV--4.13~eV of electrochromic tungsten oxide on ITO coated glass by e-beam evaporation.\cite{von1997optical} I.Valyukh \textit{et al.} fabricated  tungsten oxide films by reactive DC magnetron sputtering and performed ellipsometric data analysis based transmittance and reflectance measurements from 0.72~eV--4.13~eV.\cite{valyukh2010spectroscopic} A. Georg \textit{et al.} studied WO$_{x}$ films with varying crystallinity grown by thermal evaporation comparing optical constants within the range  from 0.62~eV--3.72~eV. \cite{georg1998comparison} 
D. H. Mendelsohn \textit{et al.}, reported on the refractive index and extinction coefficient using ellipsometry in the spectral range 0.5~eV--3.5~eV of polycrystalline electrochromic WO$_{3}$ films, grown by RF sputtering.\cite{mendelsohn1984ellipsometry} However, to the best of our knowledge, a wide spectral range including ultraviolet to vacuum ultraviolet dielectric function were not yet reported for WO$_{3}$ thin films using ellipsometry. 

Spectroscopic ellipsometry (SE) is a convenient non-destructive, non-contact optical characterization method, which has been widely employed to study thin films.\cite{rothen1945ellipsometer,rothen1945ellipsometer,schmidt2012optical,knaut2012situ,jarrendahl1998multiple,schubert1998generalized,schmidt2008generalized} The interaction between an incident polarized light beam and a stack of layered materials  of interest with plane parallel interfaces results in change in the polarization of the reflected or transmitted light beam.\cite{fujiwara2007spectroscopic} The complex-valued ratio, $\varrho$, taken between incident and reflected or transmitted electric field components of a monochromatic electromagnetic plane waves can be accurately measured in terms of the amplitude ratio, $\Psi$, and the phase difference, $\Delta$. For isotropic materials and for the case of reflection geometry the relation holds

\begin{equation}
\varrho =\frac { { r }_{ \textrm{p} } }{ { r }_{ \textrm{s} } } =\textrm{tan}\left( \Psi  \right) { e }^{ i\Delta  } ,
\label{SE_Relation}
\end{equation} 
where $r_{\textrm{p}}$ and $r_{\textrm{s}}$ are the Fresnel reflection coefficients for parallel and perpendicular polarized light, respectively.\cite{fujiwara2007spectroscopic, schubert2004infrared} 

In order to resolve both thickness and the complex-valued frequency dependent dielectric function of a thin film, a multiple sample analysis approach is necessary.\cite{tompkins2000spectroscopic,koster2011situ,hilfiker2008survey} In this method, the dielectric function can be obtained without use of physical model lineshape functions using a wavelength-by-wavelength analysis where data points at the same photon energy (wavelength) from multiple samples are simultaneously analyzed fitting for thicknesses as well as the real and imaginary part of the dielectric function.\cite{jarrendahl1998multiple} The necessary requirement for this method is that the dielectric functions of the material of interest in the thin film samples are identical. A second requirement is that the samples differ significantly in their thickness. This scheme can be extended to more complex sample structures, for example, by incorporation of the effects of surface and interface roughness, and by including additional model parameters such as model roughness layer thickness and porosity.\cite{heinemeyer2008uniaxial,easwarakhanthan2007forouhi}

The optical properties of an effective roughness layer can be calculated using an effective medium approximation which adds together the dielectric function of the material and air with a 50:50 ratio.\cite{fujiwara2007spectroscopic} By integrating the model roughness layer, the optical effect of a rough surface onto the ellipsometric spectra is successfully modeled. It has been demonstrated that the approach of the effective surface roughness model layer predicts accurately the mean square roughness of physically rough surfaces for very small roughness parameters against the wavelength.\cite{fang1997influence, petrik1998comparative} 

The electronic band-to-band transitions give rise to critical-point (CP) features in the dielectric function. A model dielectric function (MDF) approach can be formulated using physically meaningful lineshape functions with model parameters determined by a MDF analysis.\cite{adachi1993optical} In this work, CP contributions to the WO$_3$ dielectric function are identified and modeled with a 3D-M$_1$ Adachi function for the first CP and Gaussian broadened oscillators for higher energy features. The M$_1$ Adachi function is given by:\cite{adachi1993optical}
\begin{equation}
\varepsilon \left( E \right) =-A\left( { \chi  }^{ -2 }ln\left( 1+{ \chi  }^{ 2 } \right) \right), 
\label{CPM1relation}
\end{equation}

\noindent where $\chi ={ \left( E +iB \right)  }/{ E_\mathrm{c} }$ with $A$, $E_\mathrm{c}$, and $B$ are CP amplitude, transition energy, and broadening parameters, respectively. The Gaussian oscillator used for higher energy features is given by:\cite{mock2016anisotropy} 

\begin{equation}
\varepsilon_2(E) = A(e^{-(\frac{E-E_\mathrm{c}}{\sigma})^2}-e^{-(\frac{E+E_\mathrm{c}}{\sigma})^2}), \label{gaussiane2}
\end{equation}

\noindent where $\sigma = B/(2\sqrt{ln(2)})$ with $A$, $E_{\mathrm{c}}$, and $B$ are CP amplitude, center transition energy, and broadening parameters, respectively. The real part, $\varepsilon_1$, is then obtained from Kramers-Kronig integration\cite{meneses2006structure,mock2016anisotropy}
\begin{equation}
\varepsilon_1(E) = \frac{2}{\pi}P\int_0^\infty \frac{\xi\varepsilon_2(\xi)}{\xi^2-E^2}d\xi.\label{gaussiane1}
\end{equation}
In this study, we employ SE to characterize the optical properties of the WO$_{3}$ thin films. We utilize analysis of multiple samples using different thickness of WO$_{3}$ in order to determine the isotropic dielectric function without model lineshape assumptions. \cite{valyukh2010spectroscopic} Three films are deposited using an atomic layer deposition technique for $75$, $110$, and $150$ cycles under the same deposition conditions. SE data is collected and analyzed in the spectral range from $0.72-8.5$~eV at multiple angles of incidence. We therefore determine the dielectric function of the as-grown WO$_{3}$ as well as thicknesses and effective roughness thicknesses of each film. Atomic force microscopy (AFM) measurements are performed and resulting roughness is with the effective roughness parameter found in the SE analysis. We compare results from our critical point analysis of the as-deposited WO$_{3}$ films to the envelope of expected the band-to-band transitions calculated using density functional theory. 

\section{Experimental Details}

Thin films of WO$_{3}$ were deposited on silicon wafers by plasma-enhanced ALD (Fiji F200, Veeco CNT). The $(100)$ oriented wafers with native oxide were cut from low-doped, p-type conductive, single crystalline silicon. After sample insertion into the reactor, and prior to the main deposition processes, a 300~W oxygen plasma was applied for $300$~sec in order to remove residual surface contaminants. Subsequently, a stabilization period was implemented to let the sample reach a steady state temperature. ALD techniques using cyclic exposure to (tBuN)$_{2}$(Me$_{2}$N)$_{2}$W, H$_2$O, and oxygen plasma as described in Ref. \onlinecite{liu2011water}. 

\begin{figure}[htbp]
\centering
\includegraphics[width=\linewidth]{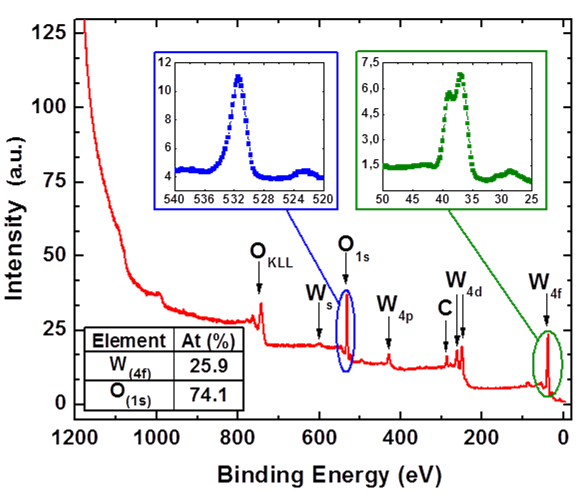}
\caption{XPS analysis of WO$_{3}$ ultra thin film fabricated with 150 ALD cycles.}
\label{XPS}
\end{figure}

In order to determine elemental and compositional information of our fabricated samples, X-ray photoelectron spectroscopy (XPS) is utilized. The resulting XPS survey spectra corresponding to the samples fabricated with 150 ALD cycles is presented in Fig. \ref{XPS}. The insets of Fig.\ref{XPS} show the spectra for O(1s) and W(4f) core levels. We find that only oxygen and tungsten are present in the film, with a chemical composition of 74.1\% and 25.9\%, respectively. These values are in good agreement with what is expected for WO$_3$ films.

SE measurements were conducted in the spectral range of 0.72--6.2~eV using a dual rotating compensator ellipsometer (RC2, J.A.~Woollam~Co.,~Inc.), and in the spectral range of 4--8.5~eV using a rotating analyzer ellipsometer with an automated compensator (VUV-VASE, J.A.~Woollam~Co.,~Inc.). All spectra were collected at angles of incidence of 45$^\circ$, 55$^\circ$, 65$^\circ$, and 75$^\circ$. All ellipsometric spectra were analyzed using ellipsometry data modeling software (WVASE32, J.A.~Woollam~Co.,~Inc.).  

\begin{figure}[htbp]
\centering
\includegraphics[width=\linewidth]{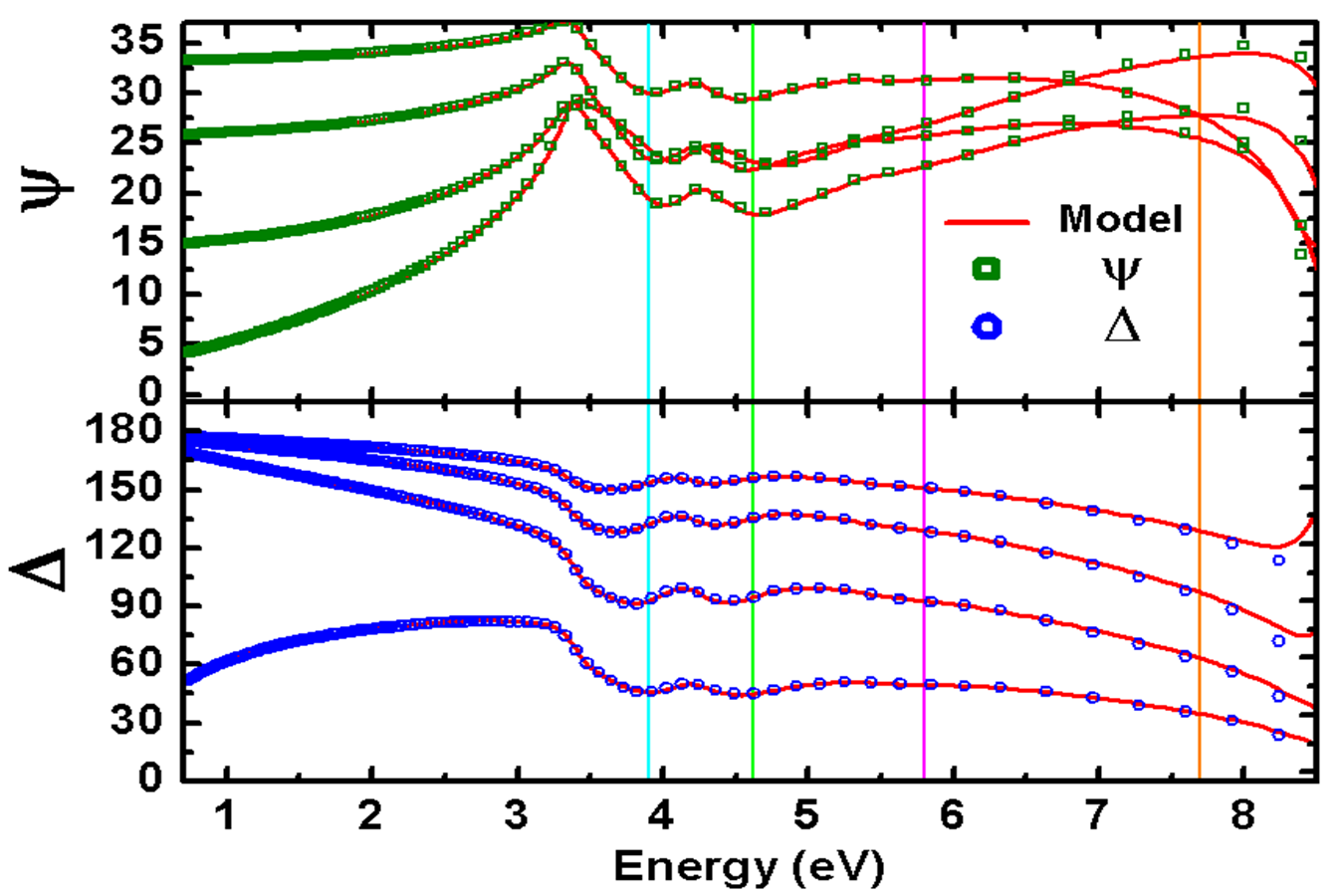}
\caption{Experimental $\Psi$ (green, squares) and $\Delta$ (blue, circles) versus photon energy for various angles of incidence collected from the WO$_3$ thin film fabricated with 150 ALD cycles on silicon substrate. MSA best-match model results (red, solid) are shown for comparison. Vertical lines indicate critical point transition energies determined by CP analysis.}
\label{psianddeltavsenergy}
\end{figure}

AFM images were collected from all samples using a multi-mode atomic force microscope (Bruker-Nanoscope III). For all measurements, the field size was chosen to be 2$\times$2 $\mu$m with a line resolution of 512$\times$512. The AFM scans were performed in tapping mode with a scan velocity of 0.1 lines per second. Image data were analyzed using Nanoscope Visualization and Analysis software. The model surface roughness parameters of the investigated samples were calculated from the image data, and obtained as R$_{q}$, the average of height deviation taken from the mean image data plane, and as R$_{a}$, the arithmetic average of the absolute values of the surface height deviations measured from the mean geometric (flat) surface plane.\cite{oraby2010atomic}

\begin{figure}[htbp]
\centering
\includegraphics[width=\linewidth]{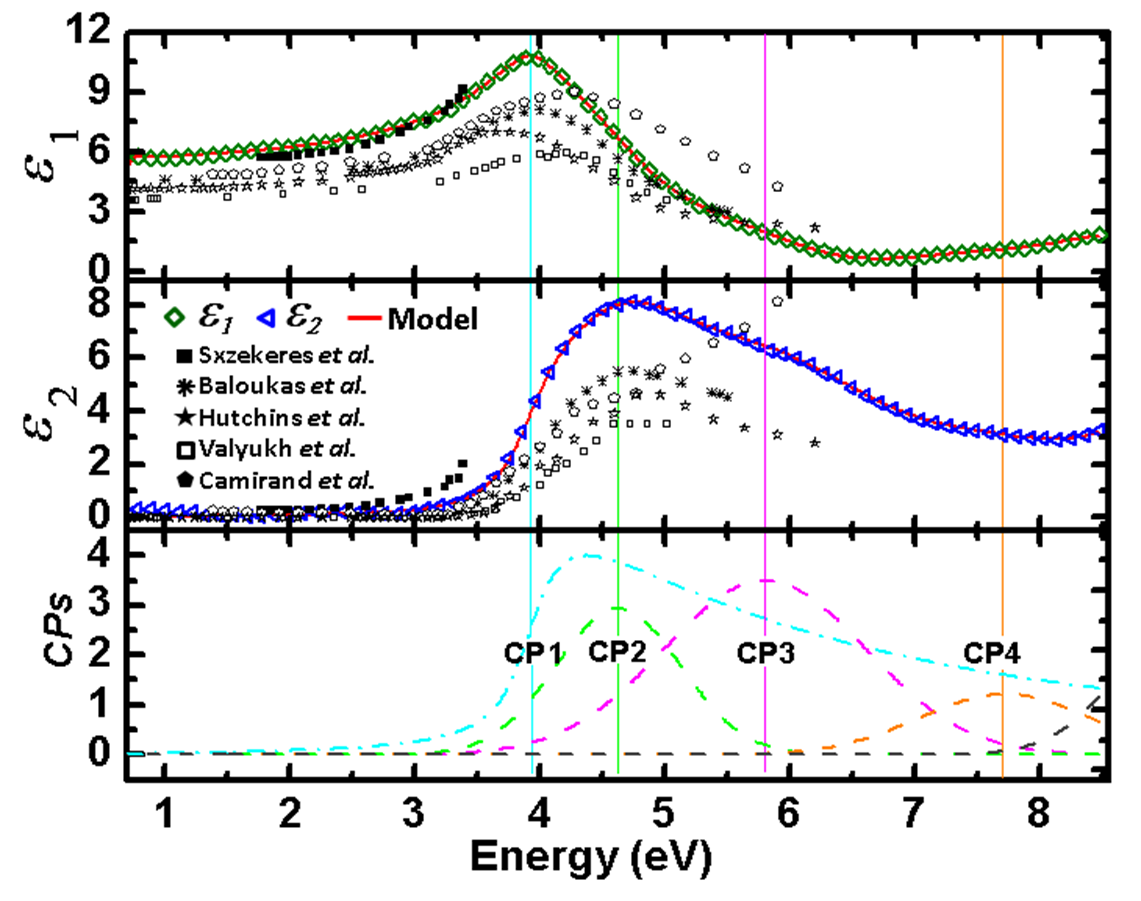}
\caption{Wavelength-by-wavelength determined dielectric constants of the WO$_{3}$ thin films extracted using the MSA approach (symbols) are shown in comparison with those from the CP analysis (red, solid). The dielectric functions of (i) CVD tungsten oxide films deposited at 400$^{o}$C (black solid square symbols) by A. Szekeresa \textit{et al}.\cite{szekeres1999optical}, (ii) thermally evaporated tungsten oxide films onto quartz substrates deposited at 350 K (black solid square symbols) by M.G. Hutchins \textit{et al}. \cite{hutchins2006structural}, (iii) magnetron sputtering of tungsten oxide films (black solid star symbols) by B. Baloukas \textit{et al}. \cite{baloukas20123}, (iv) dc magnetron sputtering of tungsten oxide films (black hollow square symbols) by I. Valyukh \textit{et al}. \cite{valyukh2010spectroscopic}, and (v) tungsten oxide films deposited via  magnetron sputtering holding the pressure at 5mTorr (black hollow square symbols) by H. Camirand \textit{et al}. \cite{camirand2015situ} are shown for comparison. The bottom section shows individual critical point contributions are all labeled except CP5 (gray, dashed line) which is centered outside the investigated spectral range. Vertical solid lines are overlaid at the E$_\textrm{c}$ energies for the CP functions.} 

\label{dielectric constant}
\end{figure}
A density functional theory approach was used for bandstructure calculations. The calculations were performed using the plane-wave density functional theory (DFT) code Quantum ESPRESSO.~\cite{[{Quantum ESPRESSO is available from http://www.qu\-an\-tum-es\-pres\-so.org. See also: }]GiannozziJPCM2009QE} Atomic coordinates and unit cell parameters were taken from Ref.~\onlinecite{Woodward1995}. We used the exchange-correlation functional of Perdew and Zunger \cite{PerdewPRB1981} and the atoms were represented by Optimized Norm-Conserving Vanderbilt (ONCV) scalar-relativistic pseudopotentials,\cite{Hamann2013} which
we generated for the PZ functional using the code ONCVPSP\cite{ONCVPSP} with the optimized parameters of the SG15 distribution of pseudopotentials.\cite{SG15} The simple monoclinic unit cell containing 8 tungsten atoms and 24 oxygen atoms was first relaxed to force levels less than 10$^{-4}$ Ry/Bohr. A regular shifted 2$\times$2$\times$2 Monkhorst-Pack grid was used for sampling of the Brillouin Zone.\cite{MonkhorstPRBGRID} A convergence threshold of 1$\times$10$^{-11}$ was used to reach self consistency with a large electronic wavefunction cut-off of 100 Ry. For the relaxed structure an additional non-scf calculation was performed on a non-shifted ($\Gamma$-centered) 2$\times$2$\times$2 grid, with additional 40 unoccupied bands and somewhat relaxed convergence criteria of 1$\times$10$^{-9}$. 

The allowed band-to-band transitions were identified by analyzing the matrix elements $|M_{cv}|^2$ of the momentum operator between conduction and valence bands at the $\Gamma$ point (Figure~\ref{DFT}). Only the transitions with significant values of the matrix elements were considered, i.e., transitions with $|M_{cv}|^2<$0.05 were discarded. Also, due to the fact that DFT underestimates the energy gap between occupied and unoccupied states a rigid shift of energy was applied, i.e., the energy of all the transitions were shifted by 2.01~eV, the value chosen to match the lowest vertical transition with first CP experimentally determined in this work. 

\section{Results and Discussion}

Spectroscopic ellipsometric data, $\Psi$ and $\Delta$ from a WO$_3$ thin film fabricated with 150 ALD cycles on silicon substrate are presented in Fig. \ref{psianddeltavsenergy} along with the corresponding best match model calculations from the multiple sample analysis. Similar data was obtained for each of the fabricated samples allowing for the extraction of the dielectric function. 

\begin{table}[ht]
\centering
\caption{Thickness parameter results from ellipsometric MSA approach. Parenthesis correspond to the 90\% confidence interval obtained from the numerical best-match data analysis.} 

\begin{tabular}{lccc}
\hline\hline
\multirow{2}{*}{Thickness (nm)} &Sample I &Sample II &Sample III  \\
&(75 cycles) & (110 cycles) &(150 cycles) \\
\hline
$d_\mathrm{WO_{3}}$    & $6.57(1)$    &  $10.52(2)$   & $14.51(1)$ \\
$d_\mathrm{R}$     &$1.48(1)$    &  $2.09(1)$   & $2.38(1)$\\
\hline\hline
\end{tabular}

\label{tabledata}
\end{table} 

The WO$_3$ layer thicknesses of the investigated samples determined by this MSA approach are presented in Table~\ref{tabledata}. The real and imaginary parts of the spectrally dependent dielectric function, $\varepsilon_1$ and $\varepsilon_2$, determined using a wavelength-by-wavelength regression analysis and are shown as green squares in Fig.~\ref{dielectric constant} together with dielectric function data from selected literature are reproduced with permission from references \onlinecite{szekeres1999optical,camirand2015situ,valyukh2010spectroscopic, baloukas20123, hutchins2006structural}. We find that optical constants are highly dependent on the deposition conditions. Additionally, surface roughness plays a large part in the determination of optical constants in the visible-vacuum ultraviolet spectral region of thin films. \cite{tompkins2015spectroscopic} 
Many of the previous studies onto the dielectric function of WO$_3$ thin films did not take the surface roughness into account\cite{garde2016gas,baloukas20123,hutchins2006structural,szekeres1999optical,ping2013optical,sun2004appearance}, which could explain some of the variation seen in Fig. \ref{dielectric constant}.

\begin{table}[ht]
\centering
\caption{The list of tungsten oxide band gap energy values which is obtained by using different methods.} 
\begin{tabular}{lccccc}
\hline \hline
 $$& Fabrication & Method & E$_{gap}$  (eV)  \\ 
\hline
Our work &ALD& Spectroscopic  & $3.93(1)$ \\
$$ &$$&  ellipsometry & $$ \\
Ref. \onlinecite{valyukh2010spectroscopic} &reactive DC& Spectroscopic  & 3.15 \\
$$ & magnetron sputtering &  ellipsometry & $$ \\
Ref. \onlinecite{szekeres1999optical} &CVD &Absorption &3.25-3.4\\
$$ &$$& spectra& $$ \\
Ref. \onlinecite{hutchins2006structural}  &Thermal & Spectrophotometric  & $3.28$ \\
$$  &evaporation & transmissivity & $$ \\
Ref. \onlinecite{baloukas20123} &Reactive RF& Spectroscopic & $3.1(1)$ \\
$$ &magnetron sputtering  & ellipsometry & $$ \\
Ref. \onlinecite{madhavi2014structural} &Reactive RF & Optical & 3.08-3.48 \\
$$ &magnetron sputtering  & transmittance &$$ \\
Ref. \onlinecite{garde2016gas} &Vacuum  & Absorption  & 3.9 \\
$$ & evaporation &  spectra & $$ \\
Ref. \onlinecite{ping2013optical} &- & DFT Calculations& 3.26\footnote{ G$_{0}$W$_{0}$ band gap calculations of $\gamma$-WO$_{3}$ are performed by considering the experimental geometry.} \\
Ref. \onlinecite{sun2004appearance} &- & DFT Calculations& 3.46\footnote{This is the energy gap of (WO$_{3}$)$_{n}$ where $n=4$ cluster.} \\
\hline \hline
\end{tabular}
\label{WO3BANDGAPS}
\end{table} 

Five line-functions were fit to the experimentally obtained dielectric constants. A single Adachi M1 function (eq. \ref{CPM1relation}) was used to render the shape from the lowest critical point feature and four Gaussian functions (eq. \ref{gaussiane2}) were used to model higher energy transitions.

The resulting dielectric function lineshapes from the critical point analysis are shown in Fig. \ref{dielectric constant} as red solid lines. Individual contributions to the dielectric function are shown in $\varepsilon_2$. Vertical lines at the determined center energies in both Figs. \ref{psianddeltavsenergy} and \ref{dielectric constant}. Best match model parameters for CP analysis lineshape functions are presented in Table \ref{CPA_table}. 

The wide variation in reported band gap energy can also be attributed to variation in deposition methods and conditions with vacuum evaporation seeming to yield the widest band gap. The energy band gap values of selected tungsten oxide thin films which are deposited and optically characterized by different methods are listed in Table \ref{WO3BANDGAPS}. Unlike the other studies \cite{garde2016gas,szekeres1999optical}, in this report, having extended spectral range unraveled four higher energy CPs in the dielectric function of our ALD fabricated WO$_{3}$ ultra thin films.

\begin{table}[ht]
\centering
\caption{The CP parameter results from the dielectric model function analysis are presented. The digits in parenthesis refer to the 90$\%$ confidence interval from the numerical best-match model regression analysis.} 
\begin{tabular}{lccccc}
\hline \hline
{}& \multicolumn{5}{c}{Critical Point Analysis} \\
\cline{2-6}
Parameter & CP1\footnote{CP1 modeled by Adachi function (CPM1), Eq.\ref{CPM1relation}.} & CP2\footnote{CP2, CP3, and CP4, CP5 modeled by Gaussian functions, Eq.\ref{gaussiane2} and Eq.\ref{gaussiane1}.} & CP3$^\textrm{b}$ & CP4$^\textrm{b}$ & CP5$^\textrm{b,}$\footnote{Transition outside investigated region determined with limited sensitivity.} \\ 
\hline
$B$ (eV) & $0.23(1)$ & $1.2(2)$ & $1.9(3)$ & $1.6(1)$ & $1.39(0)$ \\
$A$ (eV) & $1.9(7)$ & $2.93(2)$ & $3.5(2)$ & $1.22(8)$ & $3.44(4)$ \\
$E_\mathrm{c}$ (eV) & $3.93(1)$ & $4.63(6)$ & $5.80(5)$ & $7.7(4)$ & $9.3(2)$ \\
\\
\hline \hline
\end{tabular}
\label{CPA_table}
\end{table} 

\begin{figure}[htbp]
\centering
\includegraphics[width=\linewidth]{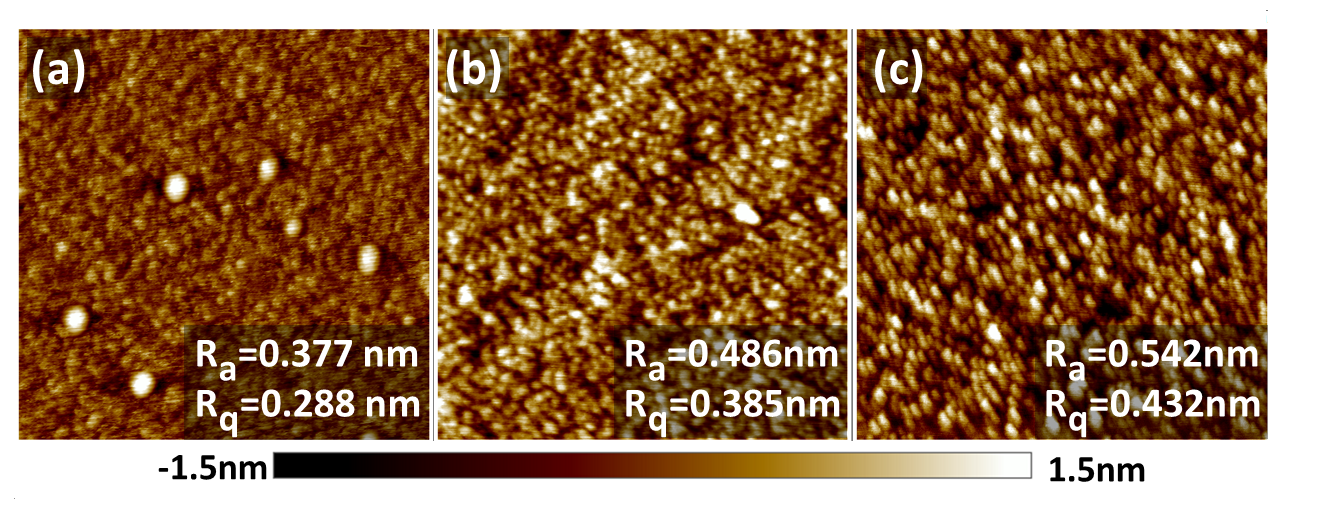}
\caption{AFM images of fabricated samples using (a) 75 ALD cycles, (b) 110 ALD cycles and (c) 150 ALD cycles.}
\label{fig:AFMimage}
\end{figure}

Increased thickness of the optical effective roughness layer is observed with increased WO$_3$ thin film thickness with determined values given in Table \ref{tabledata}. An increase of approximately 160\% is observed between the samples with 75 and 150 ALD cycles. A similar trend is observed in measured surface roughness obtained from AFM image analysis shown in Fig. \ref{fig:AFMimage}. The values for R$_a$ and R$_q$ are inlaid for each corresponding sample. An increase of approximately 140\% is seen in the R$_a$ average and approximately 150\% in the R$_q$ average between the samples with 75 and 150 ALD cycles showing excellent agreement with ellipsometric results. 

\begin{figure}[htbp]
\centering
\includegraphics[width=\linewidth]{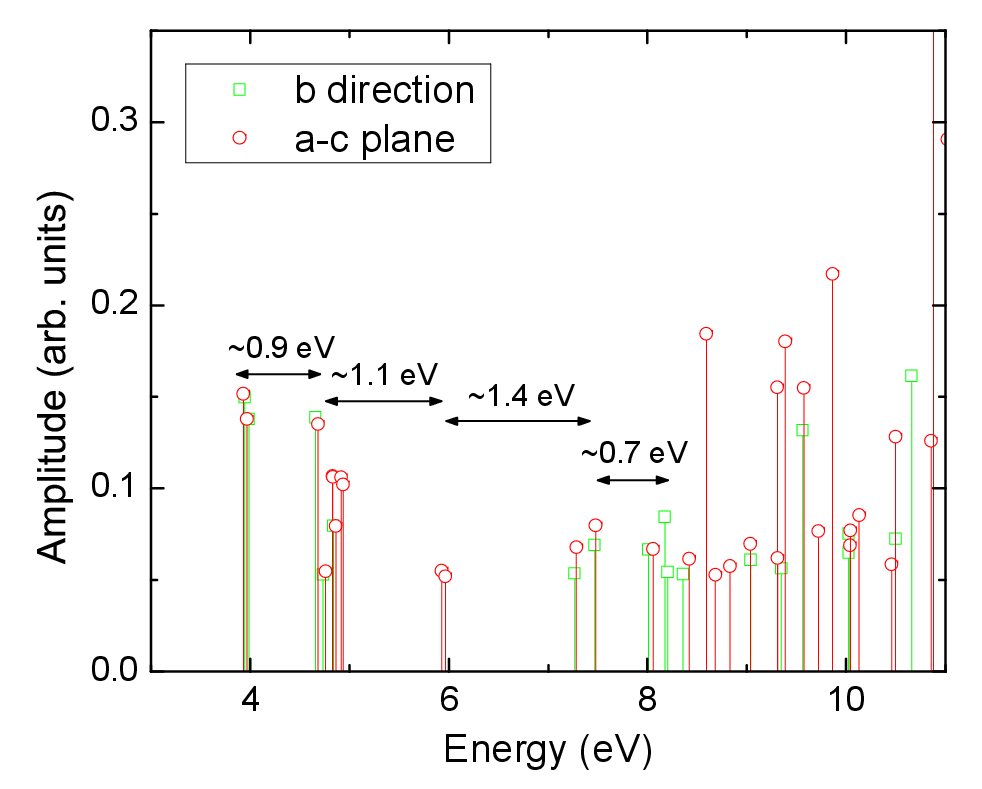}
\caption{Expected vertical electronic transitions for monoclinic WO$_3$ as obtained from DFT calculations. The transitions polarized along the symmetry axis and within the monoclinic plane are presented separately. As described in the text rigid shift was applied to align first calculated transition with the first ellipsometry-observed CP.}
\label{DFT}
\end{figure}

A snapshot of allowed optical transitions in WO$_3$ was calculated for the most common, monoclinic form of WO$_3$ and is shown in Fig. \ref{DFT}. All significant allowed transitions regardless of symmetry are plotted. Transitions are observed to fall into several clusters which are spaced in a similar way to the identified broad CP transitions from the ellipsometric analysis. 

\section{CONCLUSIONS}

Deposition of WO$_3$ by ALD was performed and spectroscopic ellipsometry was utilized to determine optical properties of the thin films. WO$_3$ thin films with various thicknesses (6.57~nm, 10.52~nm, and 14.51~nm for 75, 110, and 150~cycles, respectively) were investigated using a multiple sample analysis which allowed the thicknesses and optical properties to be decoupled. A comprehensive optical characterization using a critical point analysis was conducted in the wide spectral range of $0.72$-$8.5$~eV and compared with density functional theory calculations. Energetic locations for clusters of DFT calculated allowed transitions agree well with ellipsometric determined energy parameters of critical point features in the dielectric function.  

\section{Acknowledgements}

This work was supported in part by the National Science Foundation (NSF) through the Center for Nanohybrid Functional Materials (EPS-1004094), the Nebraska Materials Research Science and Engineering Center (MRSEC) (DMR-1420645) and awards CMMI 1337856 and EAR 1521428. The authors further acknowledge financial support by the University of Nebraska-Lincoln, the J.~A.~Woollam Co., Inc., and the J.~A.~Woollam Foundation.

\section{References}
%

\end{document}